\documentclass[%
 reprint,
superscriptaddress,
nofootinbib,
 amsmath,amssymb,
 aps,
]{revtex4-2}

\usepackage{graphicx}
\usepackage{dcolumn}
\usepackage{bm}
\usepackage{subcaption}
\usepackage{caption}
\usepackage{relsize} 

\usepackage{alphalph} 

\usepackage[left]{lineno} 

\newcommand {\KpimuePiM} {$K^{+} \rightarrow \pi^{-} \mu^{+} e^{+}$ }
\newcommand {\KpimueMuM} {$K^{+} \rightarrow \pi^{+} \mu^{-} e^{+}$ }
\newcommand {\KpimueElM} {$K^{+} \rightarrow \pi^{+} \mu^{+} e^{-}$ }

\newcommand\T{\rule{0pt}{2.2ex}}       % Top strut (2.6ex)
\newcommand\B{\rule[-1.0ex]{0pt}{0pt}} % Bottom strut (-1.2ex)

\usepackage[bottom,flushmargin,hang,multiple]{footmisc}

\begin{document}

%\preprint{APS/123-QED} %% ?????

\title{ 
% --- for CERN/arXiv preprint --- 
{\normalfont{\Large EUROPEAN ORGANIZATION FOR NUCLEAR RESEARCH} \vspace{3mm} 
\begin{flushright}
% reserved for CERN preprint number when available \\ 
CERN-090-2021
\hspace{20pt}\\12 May 2021
%\em{ To be submitted to PRL} % for preprint
\end{flushright}
}
\vspace{10mm}
% ---
Search for lepton number and flavour violation in {\boldmath $K^{+}$} and {\boldmath $\pi^{0}$} decays
}

\author{The NA62 Collaboration}
\affiliation{Authors are listed at the end of this Letter.}

%\date{12 May 2021} %\today 

\begin{abstract} 
Searches for the lepton number violating \KpimuePiM decay and the lepton flavour violating \KpimueMuM and $\pi^{0} \rightarrow \mu^{-} e^{+}$ decays are reported using data collected by the NA62 experiment at CERN in $2017$--$2018$. No evidence for these decays is found and upper limits of the branching ratios are obtained at 90\% confidence level: $\mathcal{B}(K^{+}\rightarrow\pi^{-}\mu^{+}e^{+})<4.2\times 10^{-11}$, $\mathcal{B}(K^{+}\rightarrow\pi^{+}\mu^{-}e^{+})<6.6\times10^{-11}$ and $\mathcal{B}(\pi^{0}\rightarrow\mu^{-}e^{+})<3.2\times 10^{-10}$. These results improve by one order of magnitude over previous results for these decay modes.
\end{abstract}

\maketitle

\section{Introduction}
\label{sec:Intro}
Discovery of Lepton Number (LN) or Lepton Flavour number (LF) violation would be a clear indication of new physics; although they are conserved quantum numbers in the Standard Model (SM), their conservation is not imposed by any local gauge symmetry.
Observation of neutrino oscillations provided the first proof of the non-conservation of LF, however no evidence of LN violation has been observed so far.
New physics models which explain experimental observations, such as neutrino oscillations or the possible flavour anomalies in $B$-physics~\cite{BAnomalies}, can introduce LN and LF violation.
The see-saw mechanism \cite{SeeSaw} provides a source of LN violation through the exchange of Majorana neutrinos, as in neutrinoless double beta decay.
Processes violating LF conservation can occur via the exchange of leptoquarks~\cite{Leptoquark,PS3}, of a $Z^{\prime}$ boson \cite{NewZBoson_Kaon,NewZBoson} or in SM extensions with light pseudoscalar bosons~\cite{ALP}.
Searches for kaon decays violating LN and LF conservation are powerful probes of models beyond the SM at mass scales up to $\mathcal{O}(100\,\text{TeV})$. These complement searches in $B$ meson or lepton decays, such as those producing recent limits on branching ratios $\mathcal{B}(B^{+}\rightarrow K^{+}\mu^{-}e^{+})<7.0\times10^{-9}$~\cite{LHCb19} and $\mathcal{B}(\mu^{+}\rightarrow e^{+}\gamma)<4.2\times10^{-13}$~\cite{MEG16}, which explore different aspects of new physics models.
An indirect upper limit on $\mathcal{B}(K^{+}\rightarrow\pi^{-}\mu^{+}e^{+})$ of a few units $\times10^{-11}$ has been derived from an upper limit on the $\mu^{-} + (Z,A) \to e^{+} + (Z-2,A)$ conversion probability~\cite{LittenburgShrock00}.
Previous experimental limits on LN and LF violating $K^{+}$ and $\pi^{0}$ decays are reported in Table~\ref{tab:limits}.

\begin{table}
    \caption{Summary of previous experimental limits at 90\% CL on the branching ratios of LN and LF violating $K^{+}$ and $\pi^{0}$ decays.}
    \centerline{
    %\resizebox{1.0\textwidth}{!}{
        \begin{tabular}{|l|c|}
            \cline{2-2}
                \multicolumn{1}{c|}{} &\multicolumn{1}{c|}{ Limit at 90\% CL}\T \B\\
            \hline
            $K^{+}\rightarrow\pi^{-}\mu^{+}\mu^{+}$ 
                & $<4.2 \times 10^{-11}$ (NA62 at CERN~\cite{KpillLimit})\T \B\\
            $K^{+}\rightarrow\pi^{-}e^{+}e^{+}$ 
                & $<2.2 \times 10^{-10}$ (NA62 at CERN~\cite{KpillLimit})\T \B\\
                $K^{+}\rightarrow\pi^{-}\mu^{+}e^{+}$ 
                & $<5.0 \times 10^{-10}$~~~~(E865 at BNL~\cite{KpimueMuMPiMLimit})\T \B\\
                $K^{+}\rightarrow\pi^{+}\mu^{-}e^{+}$ 
                & $<5.2 \times 10^{-10}$~~~~(E865 at BNL~\cite{KpimueMuMPiMLimit})\T \B\\
                $K^{+}\rightarrow\pi^{+}\mu^{+}e^{-}$ 
                & $<1.3 \times 10^{-11}$~~~~(E865 at BNL~\cite{KpimueElMLimit})\T \B\\
                $\pi^{0}\rightarrow\mu^{-}e^{+}$ 
                & $<3.4 \times 10^{-9}$~~~~~(E865 at BNL~\cite{KpimueMuMPiMLimit})\T \B\\
                $\pi^{0}\rightarrow\mu^{+}e^{-}$ 
                & $<3.8 \times 10^{-10}$~~~~(E865 at BNL~\cite{PiZeroElMLimit})\T \B\\
                $\pi^{0}\rightarrow\mu^{\pm}e^{\mp}$ 
                & $<3.6 \times 10^{-10}$ (KTeV at FNAL~\cite{KTeVpi0})\T \B\\
            \hline
        \end{tabular}
    }
    \label{tab:limits}
\end{table}

In this letter searches are presented for the LN violating \KpimuePiM decay ($\pi^{-}$ channel), and the LF violating decays \KpimueMuM ($\mu^{-}$ channel) and $\pi^{0}\rightarrow\mu^{-}e^{+}$, using the data collected by the NA62 experiment at the CERN SPS in $2017$--$2018$.

\section{Beamline and detector}

\begin{figure*}
        \vspace{-10pt}
        \centering
        \includegraphics[width=0.8\textwidth]{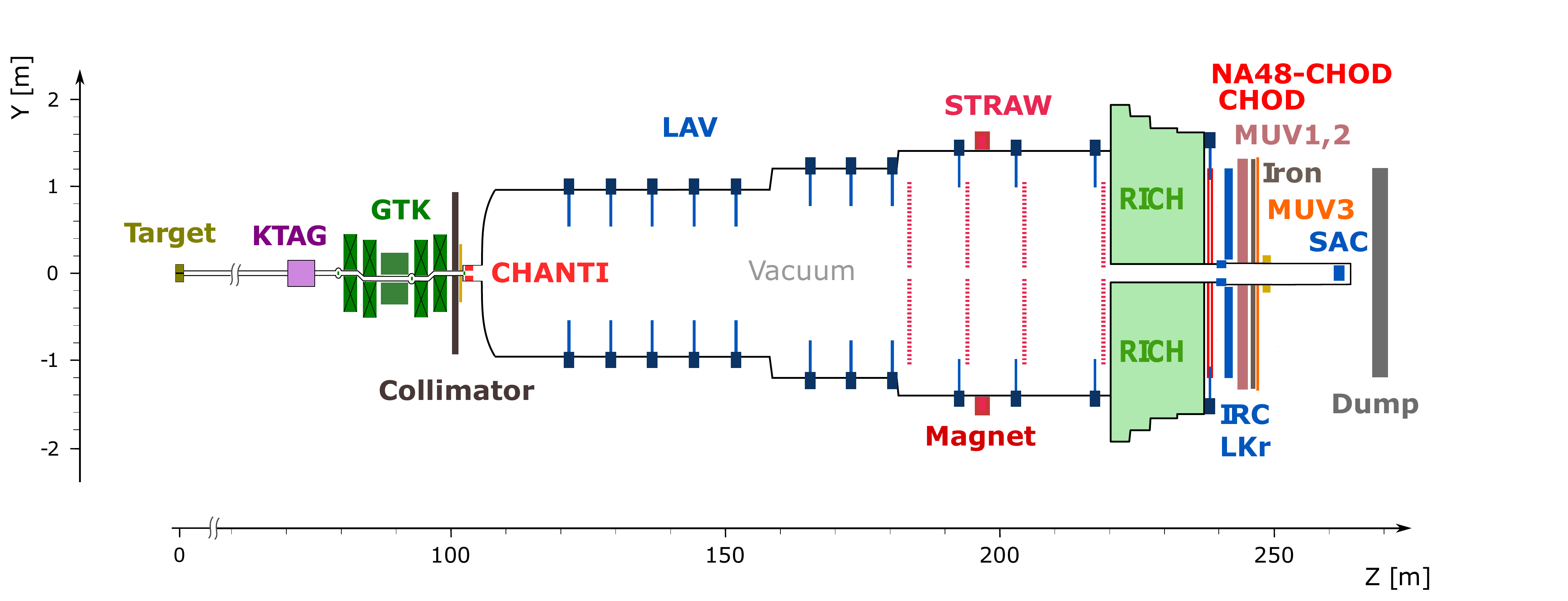}
        \vspace{-10pt}
        \caption{
        Schematic side view of the NA62 beamline and detector. Information from the CHANTI, IRC and SAC veto detectors, MUV1,2 hadronic calorimeters, and GTK beam spectrometer 
        is not used in this analysis. 
        }
        \label{fig:NA62diagram}
\end{figure*}

A sketch of the NA62 beamline and detector is shown in Figure~\ref{fig:NA62diagram} and a detailed description can be found in~\cite{NA62_DetectorPaper}. Kaons are produced in the interaction of a high intensity $400$ GeV proton beam extracted from the CERN SPS with a beryllium target.    
The resulting secondary hadron beam of positively charged particles consists of 70\% $\pi^{+}$,  23\% protons,  and 6\% $K^{+}$, with a nominal momentum of $75\,\text{GeV/}c$ ($1\%$ rms momentum bite).
Beam kaons are identified by a differential Cherenkov counter (KTAG) with $70\,\text{ps}$ time resolution and reconstructed using a silicon pixel beam spectrometer (GTK). 
The momenta and directions of charged particles produced in $K^{+}$ decays in a $75\,\text{m}$ long fiducial volume (FV) are measured by a magnetic spectrometer (STRAW). Particle identification is provided by a ring-imaging Cherenkov detector (RICH), a quasi-homogeneous liquid krypton electromagnetic calorimeter (LKr), hadronic calorimeters (MUV1,2) and a muon detector (MUV3). 
A photon veto system includes the LKr, twelve ring-shaped lead-glass detectors (LAV1--12) and small angle calorimeters (IRC and SAC).
The RICH provides a trigger time with $70\,\text{ps}$ precision. 
Two scintillator hodoscopes, NA48-CHOD and CHOD, each arranged in four quadrants, provide trigger signals and time measurements for charged particles with $200\,\text{ps}$ and $800\,\text{ps}$ precision, respectively.

\section{Data sample and trigger}
\label{sec:DataSetAndTrig}
The data sample consists of $8.3\times10^{5}$ SPS spills collected in 2017 and 2018 with a typical primary beam intensity of $2.2\times10^{12}$ protons per spill of three seconds effective duration, corresponding to a mean $K^{+}$ decay rate in the FV of 3.7 MHz.
The trigger system is composed of a hardware level (L0) and a software level (L1), with maximum output rates of 1 MHz and 10 kHz, respectively~\cite{NA62_Trigger}. The three trigger chains used for this analysis run concurrently with the trigger chain dedicated to the main goal of the experiment, the measurement of the $K^{+} \rightarrow \pi^{+} \nu \bar{\nu}$ branching ratio~\cite{NA62_PNN}: the multi-track (MT), electron multi-track ($e$MT), and muon multi-track ($\mu$MT) triggers.

The MT L0 trigger requires a signal in the RICH, and a time coincidence of signals in two opposite CHOD quadrants.
The $e$MT trigger collects a sample enriched with electrons, which deposit almost all of their energy in the LKr, by additionally requiring a minimum energy deposit of 20 GeV in the LKr (LKr20 signal). The $\mu$MT trigger selects at least one muon in the final state, requiring in addition to the MT conditions a coincident signal in the MUV3 and a minimum energy deposit of 10 GeV in the LKr (LKr10 signal). 
The common L1 trigger conditions select events with a $K^{+}$ identified by the KTAG within $5\,\text{ns}$ of the trigger time, and a track of a negatively charged particle reconstructed in the STRAW. For most of the data sample the L1 $\mu$MT trigger also requires fewer than 3 signals in total in LAV stations 2--11 within $6\,\text{ns}$ of the trigger time. 
The MT, $\mu$MT, and $e$MT trigger chains are downscaled typically by factors $D_{\text{MT}} = 100$, $D_{\mu\text{MT}} = 8$, and $D_{e\text{MT}} = 8$, respectively, but these values were varied during data-taking.

Data collected with a minimum bias trigger, requiring the presence of a signal in the NA48-CHOD at L0 and downscaled by a factor
of $400$, are used for particle identification and trigger efficiency studies.

\section{Analysis strategy and event selection}
\label{sec:Analysis}
The branching ratios for signal decays are measured relative to the normalisation channel $K^{+} \rightarrow \pi^{+} \pi^{+} \pi^{-}$ ($K_{3\pi}$) which, because of a similar topology to the signal decays, allows a first order cancellation of systematic effects related to trigger conditions and detector inefficiencies. 

The MT, $e$MT, and $\mu$MT trigger chains are used to collect signal events, and the MT trigger chain is used to collect normalisation events.

Acceptances for the signal and normalisation channels are evaluated using Monte Carlo (MC) detector simulations based on the GEANT4 toolkit~\cite{GEANT4}.

\label{sec:Selection}
The event selection identifies events comprising three tracks which point to the active region of the downstream detectors used in the analysis, are within 5~ns of the trigger time, and form a vertex of total charge +1 with a longitudinal distance from the target $105 < Z_{vtx} < 180$~m.
A vertex time is defined as the weighted mean of the track times, with weights assigned based on the time resolution of the detector (CHOD or NA48-CHOD) used to define the track time. To confirm that the beam particle is a $K^{+}$, a KTAG signal must be present within 3 ns of the vertex time.
Events with LAV signals within $3$~ns of the trigger time are rejected, providing a photon veto. 
The total three-momentum at the vertex must have a magnitude consistent with the measured mean $K^{+}$ beam momentum within $2.5\,\text{GeV}/c$ and its transverse component with respect to the beam axis is required to be less than $35\,\text{MeV/}c$, to reject events with missing energy.

For the normalisation channel selection, the three-track invariant mass reconstructed under the $3\pi$ mass hypothesis is required to be consistent with the charged kaon mass within $3\sigma_{3\pi}$, where the measured mass resolution is $\sigma_{3\pi} = 0.9~\text{MeV}/c^{2}$. 

Signal selection requires particle identification (PID) conditions using information from the LKr and MUV3 detectors to isolate candidate $\pi^{\mp} \mu^{\pm}e^{+}$ final states.
For each track the ratio, $E/p$, is calculated from the energy ($E$) of the associated LKr cluster and its momentum ($p$). 
If no signal in MUV3 is associated with the track, a pion is identified if $E/p<0.9$ while a positron is identified if $0.95<E/p<1.05$. For a positron, exactly one associated LKr cluster must be found. 
A muon is identified if a MUV3 signal is associated with the track and $E/p<0.2$. 
The range of the vertex longitudinal position is optimised to reduce the background from $K^{+}$ decays upstream of the FV.
It is required that $Z_{vtx}>107\,(111)\,\text{m}$ for the $\pi^{-}$ ($\mu^{-}$) channel.

For the $\pi^{-}$ channel selection, the mass of the $\pi^{-}e^{+}$ pair calculated under the $e^{-}e^{+}$ mass hypothesis is required to exceed $140\,\text{MeV}/c^{2}$.
This condition rejects backgrounds from $K^{+}\rightarrow\pi^{+} \pi^{0}$ and $K^{+}\rightarrow \pi^{0}\ell^{+}\nu_{\ell}$ ($\ell=\mu,e$) decays followed by $\pi^{0} \rightarrow e^{+}e^{-}\gamma$, with an $e^{-}$ misidentified as a $\pi^{-}$. 

The kinematic variable used to distinguish between signal and background is the invariant mass of the three charged tracks, $m_{\pi\mu e}$, computed by assigning the $\pi$, $\mu$, $e$ mass hypotheses to the tracks with corresponding identities defined by the PID requirements.
The $m_{\pi\mu e}$ region close to the charged kaon mass, $m_{K}$~\cite{PDG}, $478$--$510\,\text{MeV}/c^{2}$ is kept masked to avoid bias in the selection optimisation.
This includes the signal region, $490$--$498\,\text{MeV}/c^{2}$, and $12\,\text{MeV}/c^2$ wide control regions immediately below and above the signal region (denoted CR1 and CR2 respectively), used at the final stage of the analysis to validate the background prediction.
The $m_{\pi\mu e}$ resolution, obtained from simulation, is $1.4\,\text{MeV/}c^{2}$.

The search for the decay chain $K^{+} \rightarrow \pi^{+}\pi^{0}$ followed by $\pi^{0} \rightarrow \mu^{-} e^{+}$, is performed on the sample of events passing the $\mu^{-}$ channel selection by requiring that the reconstructed mass of the $\mu e$ pair is consistent with the $\pi^{0}$ mass, $|m_{\mu e}-m_{\pi^{0}}|<2\,\text{MeV}/c^{2}$. The $m_{\mu e}$ resolution obtained from simulation is $0.4\,\text{MeV/}c^{2}$.

\section{Trigger efficiency}
\label{sec:TriggerEff}
The trigger efficiency is measured with minimum bias data. 
For the abundant normalisation $K_{3\pi}$ events the efficiency is measured directly.
On the other hand, for the signal an enriched signal-like sample is used which is selected by loosening requirements on $Z_{vtx}$ and requiring that $m_{\pi\mu e}$ is outside the masked region.
The measured efficiency of the MT trigger for normalisation events is $\varepsilon_{\text{n}}=(93.2\pm0.5)\times10^{-2}$, and the result for signal-like events is consistent with $\varepsilon_{\text{n}}$ within $1\%$. 
The main source of MT trigger inefficiency is the STRAW condition at L1, and the uncertainty accounts for variations in the measured efficiency over time.

\begin{figure}
    \centering
    \includegraphics[width=0.49\textwidth]{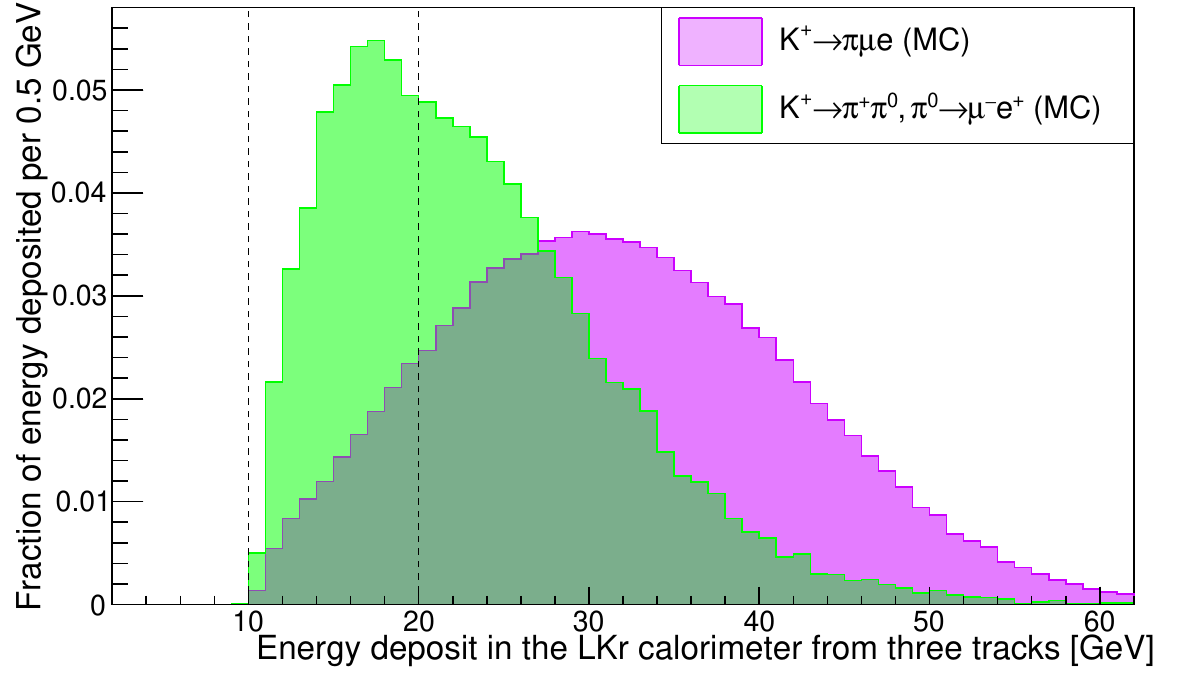}
    \caption{
    Distributions of energy deposited in the LKr associated with the three selected STRAW tracks for events passing the signal selection, $K^{+}\rightarrow\pi^{\pm}\mu^{\mp}e^{+}$ and $K^{+}\rightarrow\pi^{+}\pi^{0}$ followed by $\pi^{0}\rightarrow\mu^{-}e^{+}$,
    obtained from MC simulations after data-driven corrections to the energy of the LKr pion cluster.}
    \label{fig:ELKrSignalHistos}
\end{figure}

The L0 MUV3 and L1 LAV conditions in the $\mu$MT trigger 
have negligible inefficiency for signal-like events since similar conditions are applied offline in the selection. The efficiencies of the LKr10 and LKr20 conditions present in the $\mu$MT and $e$MT triggers, respectively, depend on the total energy deposited in the LKr.
The energy deposited by the pion in the LKr is not precisely reproduced in simulations, so a correction to this quantity is applied based on measurements. After this correction, energy-dependent trigger inefficiencies are applied in the simulation. 
The softer electron spectrum for $K^{+}\rightarrow\pi^{+}\pi^{0}$ followed by $\pi^{0}\rightarrow\mu^{-}e^{+}$ decays, with respect to $K^{+}\rightarrow\pi^{\pm}\mu^{\mp}e^{+}$ (Figure~\ref{fig:ELKrSignalHistos}), leads to a lower efficiency of the LKr10 and LKr20 trigger conditions, as will be shown below. 

\section{Normalisation to $\mathbf{K}_{\mathbf{3}\mathbf{\pi}}$ decay}
The effective number of $K^+$ decays in the FV is 
    \begin{eqnarray}
        N_{K}~&&= \mathlarger{\mathlarger{\sum\limits_{i}}}N_{K}^{i} 
            = \frac{1}{\mathcal{B}(K_{3\pi})A_{\text{n}}\varepsilon_{\text{n}}} \cdot \mathlarger{\mathlarger{\sum\limits_{i}}}\left( N_{3\pi}^{i} \frac{D_{\text{MT}}^{i}}{D_{\text{eff}}^{i}}\right) \nonumber\\
        &&= (1.33 \pm 0.02)\times10^{12}\,,
        \label{eq:Nk}
     \end{eqnarray}
where the index $i$ runs over data-taking periods defined by constant trigger downscaling factors, $N_{3\pi}^{i}$ are the numbers of normalisation $K_{3\pi}$ events selected with the MT trigger with downscaling factor $D_{\text{MT}}^{i}$, and $D_{\text{eff}}^{i}$ are the effective downscaling factors of the three signal trigger chains.  
These are evaluated as
\begin{equation}
    D_{\text{eff}}^{i}=\mkern-3mu\bigg[ 1 - \bigg(1- \frac{1}{D_{\text{MT}}^{i}}\bigg)\mkern-3mu \bigg(1- \frac{1}{D_{\mu\text{MT}}^{i}}\bigg)\mkern-3mu \bigg(1- \frac{1}{D_{e\text{MT}}^{i}}\bigg)\mkern-3mu \bigg]^{-1}
    \label{eq:Deff} 
\end{equation}
    
and vary in the range $3.2$--$6.9$.
In Eq.~(\ref{eq:Nk}), $\mathcal{B}(K_{3\pi}) = (5.583 \pm 0.024)\times 10^{-2}$~\cite{PDG} and $A_{\text{n}}=10.18\times 10^{-2}$ are the branching ratio and selection acceptance (determined using simulation) of the $K_{3\pi}$ decay, and $\varepsilon_{\text{n}}$ is the efficiency of the MT trigger for the normalisation channel.
The total number of selected $K_{3\pi}$ events collected with the MT trigger is $\sum_{i}N_{3\pi}^{i}=2.73\times10^{8}$. 
The quoted uncertainty in $N_{K}$ accounts for any inaccuracy in the description of the beam momentum spectrum and STRAW inefficiency in simulations. 

\section{Backgrounds}
\label{sec:BackgroundStudies}
Backgrounds arise from $K^{+}$ decays followed by particle misidentification and $\pi^{\pm}\rightarrow\ell^{\pm}\nu_{\ell}$ ($\ell=\mu,e$) decays in flight.
The probability of at least one $\pi^{\pm}$, from a $K_{3\pi}$ decay in the FV, to decay upstream of the LKr is found using simulations to be $7.5\%$, with the ratio of decay rates $\Gamma(\pi^{\pm}\rightarrow e^{\pm}\nu_{e})/\Gamma(\pi^{\pm}\rightarrow\mu^{\pm}\nu_{\mu})=1.23\times10^{-4}$~\cite{PDG}.

\subsection{Particle misidentification}
\label{sec:PmisID}
Misidentification of $\pi^{\pm} \rightleftharpoons e^\pm$ arises from $E/p$ measurements.
The misidentification probabilities are measured using samples of $K_{3\pi}$ ($\pi^{\pm} \rightarrow e^{\pm}$) and $K^{+}\rightarrow\pi^{+} \pi^{0}$ followed by $\pi^{0} \rightarrow e^{+}e^{-}\gamma$ ($e^{\pm} \rightarrow \pi^{\pm}$) decays, collected with the minimum bias trigger. 
In each sample, the measured contamination from other $K^{+}$ decays is below $10^{-4}$. 
The misidentification probabilities are momentum dependent with values $P(\pi^{\pm}\rightarrow e^{\pm}) = (4-5)\times10^{-3}$ and $P(e^{\pm}\rightarrow \pi^{\pm}) = (1-3)\times10^{-2}$.

Misidentification of $\pi^{\pm}$ as $\mu^{\pm}$ arises from accidental matching of tracks  with MUV3 signals or pion-induced showers in hadron calorimeters producing muons. Accidental MUV3 signals are simulated using rates measured in time sidebands within $45$--$75\,\text{ns}$ of the trigger time, and hadronic showers are simulated using GEANT4. The misidentification probability is position and momentum dependent, with values of $P(\pi^{\pm}\rightarrow \mu^{\pm})=(2-3)\times10^{-3}$.

Misidentification of $\mu^{\pm}$ as $\pi^{\pm}$ occurs due to inefficiency of the MUV3 detector. 
This inefficiency is measured to be $1.5\times10^{-3}$ using kinematically selected $K^{+}\rightarrow\mu^{+}\nu_{\mu}$ decays from minimum-bias data, and beam halo muons.

The misidentification of $e^{\pm}$ as $\mu^{\pm}$, with probability $P(e^{\pm}\rightarrow\mu^{\pm})=\mathcal{O}(10^{-8})$, occurs when an $e^{\pm}$ is absorbed or scattered inelastically upstream of the LKr.
In this case no LKr energy deposit is recorded, and the track is matched with an accidental signal in MUV3. The misidentification probability is measured from data using a sample of MUV3 signals in time sidebands and depends on track momentum and extrapolated track position at MUV3.

\subsection{Background evaluation}
Simulations that include data-driven corrections are used to predict the background. Each simulated event is assigned a weight, which accounts for misidentification probabilities and corrects for discrepancies between data and simulations in energy deposited by $\pi^{\pm}$ in the LKr, as well as in the beam momentum spectrum.

The number of selected data events with $m_{\pi\mu e}<478\,\text{MeV}/c^{2}$ agrees with predictions from simulations within $3\%$ for both the $\pi^{-}$ and $\mu^{-}$ channels (Figure~\ref{fig:Kpimue}).
The composition of backgrounds is similar in the control regions (CR1 and CR2) and in the signal regions. After unmasking the control regions, the predicted and observed numbers of events are largely consistent (Table~\ref{tab:BkgPred_CRs_Simplified}).
The predicted numbers of background events from each source in the signal regions are given in Table~\ref{tab:BkgPred_SR}.
The main contributions to the quoted uncertainties are the limited statistics of the simulations and the accuracy of the misidentification models.

\begin{table}
    \caption{
    Predicted backgrounds and observed numbers of events in control regions CR1 and CR2.
    }
    \centerline{
    %\resizebox{0.9\textwidth}{!}{
        \begin{tabular}{|l|c|c|c|c|}
            \cline{2-5}%\hline
            \multicolumn{1}{c|}{} & \multicolumn{2}{c|}{$K^{+}\rightarrow\pi^{-}\mu^{+}e^{+}$} & \multicolumn{2}{c|}{$K^{+}\rightarrow\pi^{+}\mu^{-}e^{+}$} \T \B \\
            %\cline{2-5}%\cline{1-1} 
            \multicolumn{1}{c|}{} & CR1 & CR2 & CR1 & CR2 \T \B\\  
            \hline
            %Total predicted background
            %Total bkg.
            Predicted
                & $1.68\pm0.20$ & $1.66\pm0.26$
                & $3.41\pm0.54$ & $1.27\pm0.40$ \T \B\\
            \hline
            %Numbers of observed events
            Observed
                & $2$ & $4$ & $2$ & $0$ \T \B\\
            \hline
        \end{tabular}
    }
    %}
    \label{tab:BkgPred_CRs_Simplified}
\end{table}
\begin{table}
    \caption{
    Predicted numbers of background events in signal regions. Decays upstream of the FV are the primary component of the $K^{+} \rightarrow \pi^{+}\pi^{+}\pi^{-}$ background.
    }
    \centerline{
    %\resizebox{1.0\textwidth}{!}{
        \begin{tabular}{|l|c|c|c|}
            \hline
            Source & $K^{+}\rightarrow\pi^{-}\mu^{+}e^{+}$ & $K^{+}\rightarrow\pi^{+}\mu^{-}e^{+}$ & $\pi^{0}\rightarrow\mu^{-}e^{+}$ \T \B\\
            \hline
            $K^{+}\rightarrow\pi^{+}\pi^{+}\pi^{-}$ 
                & $0.22\pm0.15$ 
                & $0.84\pm0.34$ 
                & $0.22\pm0.15$ \T \B\\
            $K^{+}\rightarrow\pi^{+}e^{+}e^{-}$ 
                & $0.63\pm0.13$ 
                & negligible 
                & negligible \T \B\\ 
            $K^{+}\rightarrow\mu^{+}\nu_{\mu}e^{+}e^{-}$ 
                & $0.13\pm0.02$ 
                & negligible 
                & negligible \T \B\\ 
            $K^{+}\rightarrow\pi^{+}\pi^{-}e^{+}\nu_{e}$ 
                & $0.07\pm0.02$ 
                & $0.05\pm0.03$ 
                & $0.01\pm0.01$ \T \B\\
            $K^{+}\rightarrow\pi^{+}\mu^{+}\mu^{-}$   
                & $0.01\pm0.01$ 
                & $0.02\pm0.01$ 
                & negligible \T \B\\
                %& $0.00533\pm0.00071$ 
                %& $0.0151\pm0.0017$ \\
            $K^{+}\rightarrow e^{+}\nu_{e}\mu^{+}\mu^{-}$ 
                & $0.01\pm0.01$ 
                & $0.01\pm0.01$ 
                & negligible \T \B\\
                %& $0.0111\pm0.0067$ 
                %& $0.0136\pm0.0061$ \\
	    \hline
            Total %background expected
                & $1.07\pm0.20$ 
                & $0.92\pm0.34$ 
                & $0.23\pm0.15$ \T \B\\
            \hline
            %Data & $0$ & $2$ \\
            %\hline
        \end{tabular}
    }
    %}
    \label{tab:BkgPred_SR}
\end{table}

\section{Single event sensitivity}

The single event sensitivities, $\mathcal{B}_{\text{SES}}^{i}$, defined for each process as the branching ratio corresponding to the observation of one signal event, are computed for each data-taking period, $i$, as
\begin{equation}
    \mathcal{B}_{\text{SES}}^{i} 
    = \frac{1}{N_{K}^{i}A_{\text{s}}\varepsilon_{\text{s}}^{i}} 
    = \mathcal{B}(K_{3\pi})\frac{A_{\text{n}}D_{\text{eff}}^{i}}{A_{\text{s}}N_{3\pi}^{i}D_{\text{MT}}^{i}}\frac{\varepsilon_{\text{n}}}{\varepsilon_{\text{s}}^{i}} \,,
    \label{eq:SESi}
\end{equation}
where $A_{\text{s}}$ are the signal acceptances (computed using simulations assuming uniform phase-space density), and $\varepsilon_{\text{s}}^{i}$ are the trigger efficiencies for signal events, which vary due to changes in trigger downscaling factors. Efficiencies for trigger components present in both normalisation and signal trigger chains cancel in Eq.~(\ref{eq:SESi}) to $1\%$ precision, except for the LKr10(20) components ($\varepsilon_{\text{LKr10(20)}}$), which depend on the energy deposited in the LKr and are not present in the MT trigger chain. Therefore %as 
\begin{equation}
    \frac{\varepsilon_{\text{s}}^{i}}{\varepsilon_{\text{n}}} \mkern-3mu =\mkern-3mu \bigg[ \mkern-1mu 1- \mkern-3mu \bigg(1- \frac{1}{D_{\text{MT}}^{i}}\bigg) \mkern-3mu \bigg(1- \frac{\varepsilon_{\text{LKr10}}}{D_{\mu\text{MT}}^{i}}\bigg) \mkern-3mu \bigg(1- \frac{\varepsilon_{\text{LKr20}}}{D_{e\text{MT}}^{i}}\bigg) \mkern-3mu\bigg] \mkern-3mu D_{\text{eff}}^{i}.
    \label{eq:effratio}
\end{equation}
A summary of inputs to the single event sensitivity calculation is given in Table ~\ref{tab:SES_}.
For the $\pi^{0}\rightarrow\mu^{-}e^{+}$ search, 
$\mathcal{B}_{\text{SES}}^{i}$ is divided by $\mathcal{B}(K^{+}\rightarrow\pi^{+}\pi^{0})=(20.67 \pm0.08)\times 10^{-2}$~\cite{PDG}.

\begin{table}
    \caption{
    Summary of inputs to the single event sensitivity calculation 
    and corresponding resulting values for each search.
    The signal acceptances, $A_{\text{s}}$, are displayed with statistical uncertainties only; 
    other uncertainties quoted are quadratic sums
    of the statistical and systematic uncertainties.
    }
    \centerline{
    %\resizebox{0.9\textwidth}{!}{
        \begin{tabular}{|c|c|c|c|}
            \hline
            % variable 
                & $K^{+}\rightarrow\pi^{-}\mu^{+}e^{+}$ 
                & $K^{+}\rightarrow\pi^{+}\mu^{-}e^{+}$  
                & $\pi^{0}\rightarrow\mu^{-}e^{+}$ \T \B\\
            \hline
            $A_{\text{s}}\times10^{2}$%\,[\%]$ 
                & $4.90\pm0.02$%_{\text{stat}}$
                %& $(4.90\pm0.02_{\text{stat}})\%$ %& $(4.90\pm0.16)\%$  %& $(4.72\pm0.15)\%$   %& $(4.90\pm0.02_{stat})\%$ 
                & $6.21\pm0.02$%_{\text{stat}}$
                %& $(6.21\pm0.02_{\text{stat}})\%$ %$(6.21\pm0.11)\%$  %& $(6.09\pm0.11)\%$   %& $(6.21\pm0.02_{stat})\%$ 
                & $3.11\pm0.02$%_{\text{stat}}$ 
                \T\B\\ 
                %& $(3.11\pm0.02_{\text{stat}})\%$ \\ %$(3.11\pm0.17)\%$ \\ %& $(3.58\pm0.19)\%$ \\ %& $(3.11\pm0.02_{stat})\%$ \\
            $\varepsilon_{\text{LKr10}}\times10^{2}$ 
                & $97.5\pm1.3\phantom{0}$ %$(97.5\pm1.3)\%$
                & $97.5\pm1.3\phantom{0}$ %$(97.5\pm1.3)\%$ 
                & $92.9\pm1.2\phantom{0}$\T\B\\ %$(92.9\pm1.2)\%$ \\
            $\varepsilon_{\text{LKr20}}\times10^{2}$ 
                & $74.1\pm1.6\phantom{0}$ %$(74.1\pm1.6)\%$
                & $73.3\pm1.6\phantom{0}$ %$(73.3\pm1.6)\%$ 
                & $45.3\pm1.0\phantom{0}$\T\B\\ %$(45.3\pm1.0)\%$ \\
            %$\varepsilon_{\text{s}}/\varepsilon_{\text{n}}$ %(84.2\pm2.1)\%$
            %    & $(87.4\pm2.7)\%$   %& $(80.6\pm)\%$
            %    & $(85.9\pm2.2)\%$   %& $(79.2\pm)\%$
            %    & $(73.2\pm3.0)\%$\\ %& $(67.5\pm)\%$ \\
	    \hline %---Brigitte suggestion
	    \hline %----Elisa proposal, 2 hlines
            $\mathcal{B}_{\text{SES}}\times10^{11}$
                & $1.82\pm0.08$%)\times10^{-11}$ 
                & $1.44\pm0.05$%)\times10^{-11}$ 
                & $13.9\pm0.9\phantom{0}$\T\B\\%)\times10^{-10}$ \\
            \hline
        \end{tabular}
    %}
    }
    \label{tab:SES_}
\end{table}

The quantity $\mathcal{B}_{\text{SES}}$ for the full data set is given by 
\begin{equation}
    \mathcal{B}_{\text{SES}} = \left[ \sum\limits_{i}(\mathcal{B}_{\text{SES}}^{i})^{-1} \right]^{-1}\,,
\end{equation}
and results are shown in Table~\ref{tab:SES_}. 
The uncertainty in $\mathcal{B}_{\text{SES}}$ includes the external error from the branching fractions $\mathcal{B}(K_{3\pi})$ and $\mathcal{B}(K^{+}\rightarrow\pi^{+}\pi^{0})$, each $0.4\%$ in relative terms, and search-specific systematic uncertainties $(2\%-7\%)$ of $\mathcal{B}_{\text{SES}}$, assigned to account for the precision of the data-driven corrections applied in simulations. 
For the LF violating \KpimueElM decay $\mathcal{B}_{\text{SES}}=(1.46\pm0.06)\times10^{-11}$ and for the $\pi^{0}\rightarrow\mu^{+}e^{-}$ decay $\mathcal{B}_{\text{SES}}=(15.9\pm1.1)\times10^{-11}$. Neither of these sensitivities are competitive with previous searches~\cite{KpimueElMLimit,KTeVpi0}.

\section{Results}
\label{sec:Results}
\begin{figure*}
    \centering
    \begin{subfigure}[b]{0.42\textwidth} 
        \includegraphics[width=1.0\textwidth]{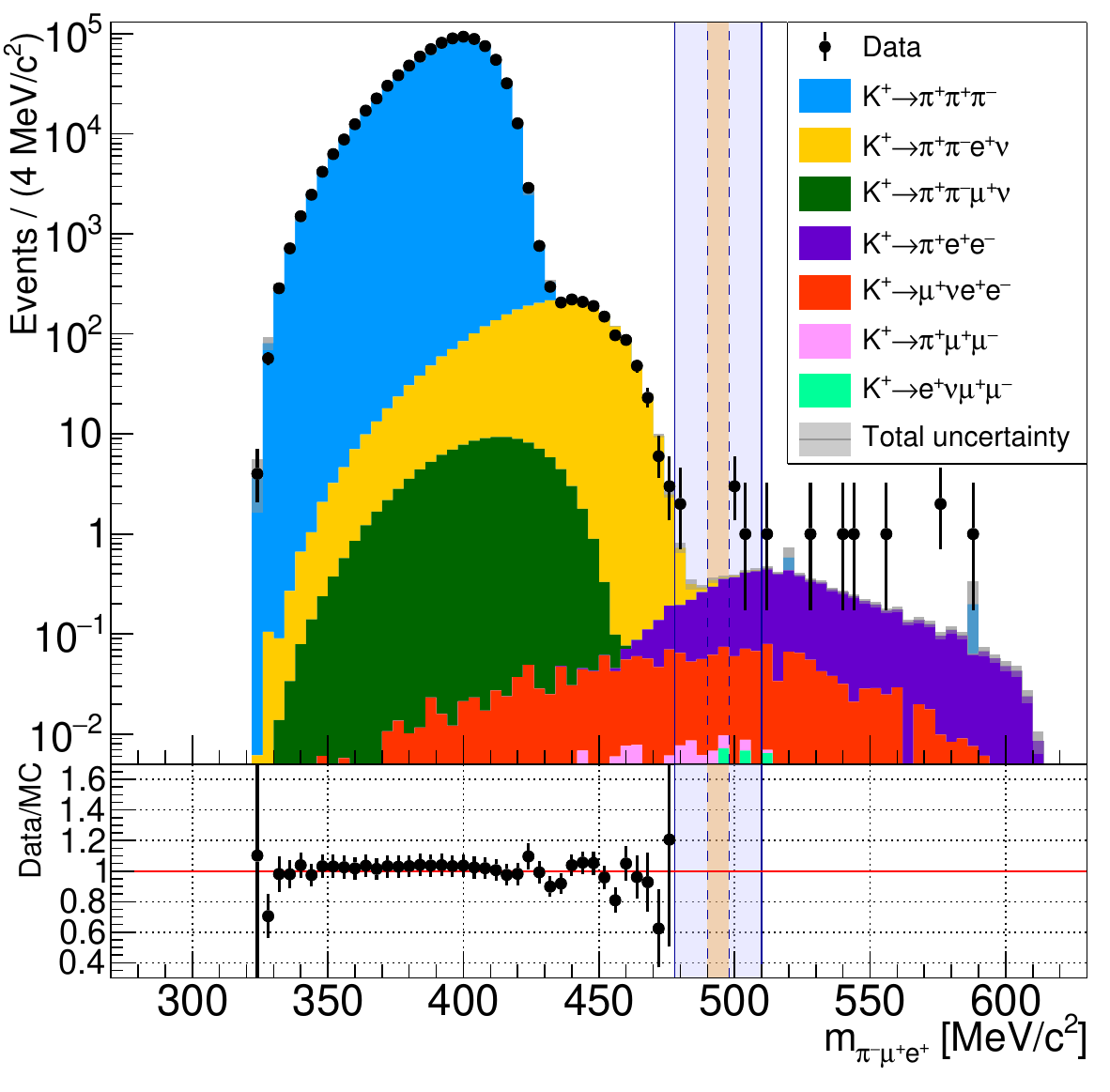}
        \label{fig:Kpimue_piM} 
    \end{subfigure}
    \begin{subfigure}[b]{0.095\textwidth}
    \hspace{1pt}
    \end{subfigure}
    \begin{subfigure}[b]{0.42\textwidth} 
        \includegraphics[width=1.0\textwidth]{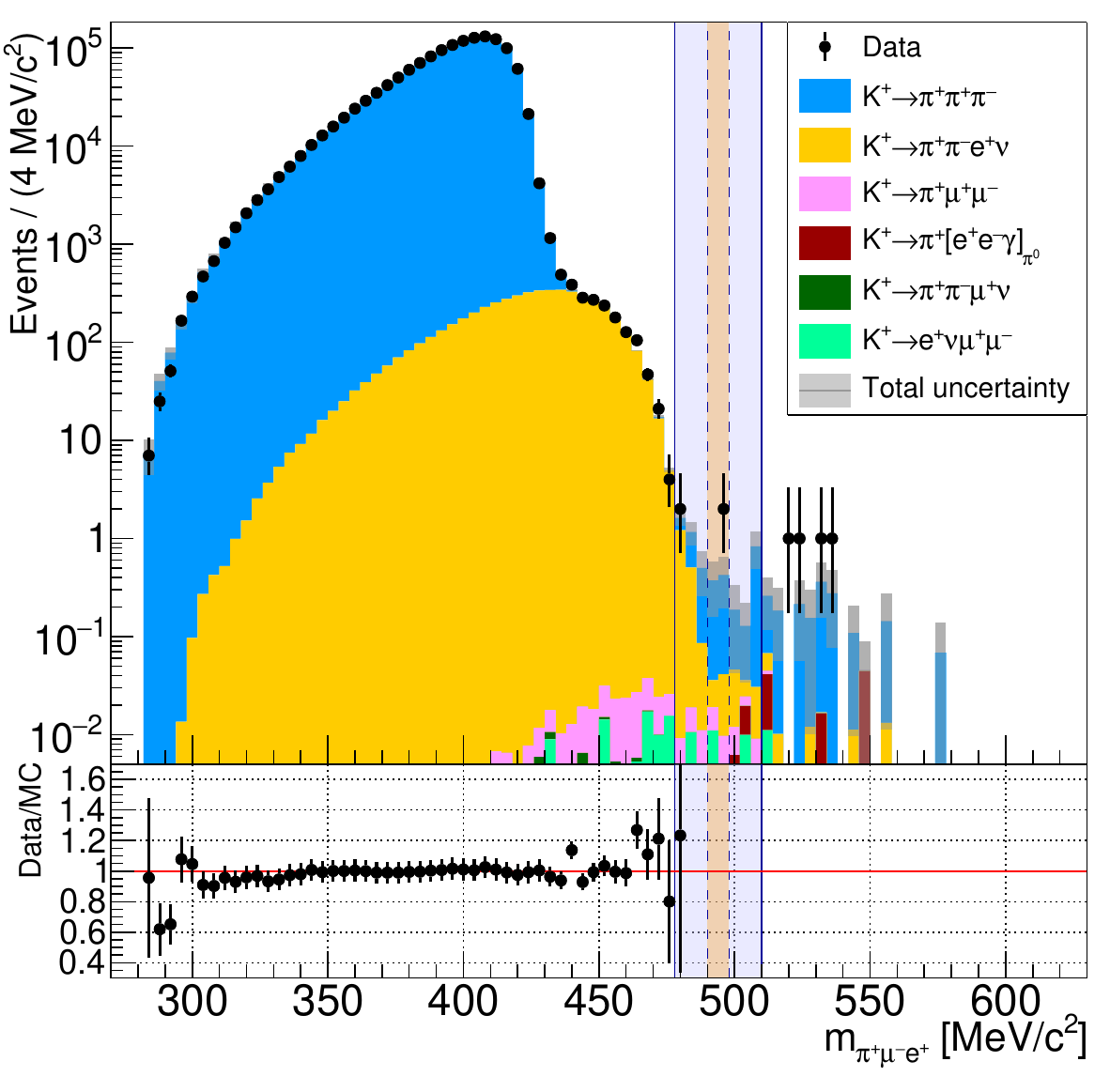}
        \label{fig:Kpimue_muM}
    \end{subfigure}
    \vspace{-15pt}
    \caption{
     Reconstructed $m_{\pi\mu e}$ spectra for selected events in searches for $K^{+}\rightarrow\pi^{-}\mu^{+}e^{+}$ (left) and $K^{+}\rightarrow\pi^{+}\mu^{-}e^{+}$ (right) for data (black markers) and simulated background (filled areas) samples. Ratios between the observed numbers of data events and the predicted numbers of events from MC simulations are shown in the lower panels.
    } 
    \label{fig:Kpimue}
\end{figure*}

After unmasking the signal regions, the mass spectra for the \KpimuePiM and \KpimueMuM searches are shown in Figure~\ref{fig:Kpimue}.
The numbers of predicted backgrounds ($n_{\text{bg}}$) and observed events ($n_{\text{obs}}$) in the signal regions are listed below
\begin{align}
    K^{+}\rightarrow\pi^{-}\mu^{+}e^{+}\,:\,\,\,\,\,n_{\text{bg}}=1.07\pm0.20\,,\,\,\,\,n_{\text{obs}}=0; \nonumber \\
    K^{+}\rightarrow\pi^{+}\mu^{-}e^{+}\,:\,\,\,\,\,n_{\text{bg}}=0.92\pm0.34\,,\,\,\,\,n_{\text{obs}}=2; \nonumber\\
    \pi^{0}\rightarrow\mu^{-}e^{+}\,\,\,:\,\,\,\,\,n_{\text{bg}}=0.23\pm0.15\,,\,\,\,\,n_{\text{obs}}=0. \nonumber
\end{align}

The observations are consistent with the background predictions, and upper limits are set for the branching ratios using the $\text{CL}_{\text{S}}$ method~\cite{Read02} with a likelihood ratio test statistic. 
The upper limits obtained at $90\%$ CL are 
\begin{align}
    \mathcal{B}(K^{+}\rightarrow\pi^{-}\mu^{+}e^{+}) & < 4.2\times10^{-11}; \nonumber\\
    \mathcal{B}(K^{+}\rightarrow\pi^{+}\mu^{-}e^{+}) & < 6.6\times10^{-11}; \nonumber\\
    \mathcal{B}(\pi^{0}\rightarrow\mu^{-}e^{+}) & < 3.2\times10^{-10}. \nonumber
\end{align}

\section{Conclusions}
\label{sec:Summary}

Searches for the LN violating $K^{+}\rightarrow\pi^{-}\mu^{+}e^{+}$, and LF violating \KpimueMuM and $\pi^{0}\rightarrow\mu^{-}e^{+}$ decays are reported.
No evidence for these decays is found and upper limits are established at 90\% confidence level:
$\mathcal{B}(K^{+}\rightarrow\pi^{-}\mu^{+}e^{+})<4.2\times10^{-11}$,  $\mathcal{B}(K^{+}\rightarrow\pi^{+}\mu^{-}e^{+})<6.6 \times10^{-11}$, and $\mathcal{B}(\pi^{0}\rightarrow\mu^{-}e^{+})<3.2\times10^{-10}$. These results improve on previous searches~\cite{KpimueMuMPiMLimit} by one order of magnitude. 
NA62 resumes data-taking in 2021, with the primary objective of improving the precision of the study of the $K^{+}\rightarrow\pi^{+}\nu\bar{\nu}$ decay~\cite{NA62_PNN}, but also with the possibility of collecting additional data to study LN and LF violating decays.

\begin{acknowledgments}
It is a pleasure to express our appreciation to the staff of the CERN laboratory and the technical
staff of the participating laboratories and universities for their efforts in the operation of the
experiment and data processing.

The cost of the experiment and its auxiliary systems was supported by the funding agencies of 
the Collaboration Institutes. We are particularly indebted to: 
F.R.S.-FNRS (Fonds de la Recherche Scientifique - FNRS), Belgium;
%BMES (Ministry of Education, Youth and Science), Bulgaria;
NSERC (Natural Sciences and Engineering Research Council), funding SAPPJ-2018-0017 Canada;
%NRC (National Research Council) contribution to TRIUMF, Canada;
MEYS (Ministry of Education, Youth and Sports),  Czech Republic;
BMBF (Bundesministerium f\"{u}r Bildung und Forschung) contracts 05H12UM5, 05H15UMCNA and 05H18UMCNA, Germany;
INFN  (Istituto Nazionale di Fisica Nucleare),  Italy;
MIUR (Ministero dell'Istruzione, dell'Universit\`a e della Ricerca),  Italy;
CONACyT  (Consejo Nacional de Ciencia y Tecnolog\'{i}a),  Mexico;
IFA (Institute of Atomic Physics) Romanian CERN-RO No.1/16.03.2016 and Nucleus Programme PN 19 06 01 04,  Romania;
INR-RAS (Institute for Nuclear Research of the Russian Academy of Sciences), Moscow, Russia; 
JINR (Joint Institute for Nuclear Research), Dubna, Russia; 
NRC (National Research Center)  ``Kurchatov Institute'' and MESRF (Ministry of Education and Science of the Russian Federation), Russia; 
MESRS  (Ministry of Education, Science, Research and Sport), Slovakia; 
CERN (European Organization for Nuclear Research), Switzerland; 
STFC (Science and Technology Facilities Council), United Kingdom;
NSF (National Science Foundation) Award Numbers 1506088 and 1806430,  U.S.A.;
ERC (European Research Council)  ``UniversaLepto'' advanced grant 268062, ``KaonLepton'' starting grant 336581, Europe.

Individuals have received support from:
Charles University Research Center (UNCE/SCI/013), Czech Republic;
Ministero dell'Istruzione, dell'Universit\`a e della Ricerca (MIUR  ``Futuro in ricerca 2012''  grant RBFR12JF2Z, Project GAP), Italy;
%Russian Foundation for Basic Research  (RFBR grants 18-32-00072, 18-32-00245), Russia; 
Russian Science Foundation (RSF 19-72-10096), Russia;
the Royal Society  (grants UF100308, UF0758946), United Kingdom;
STFC (Rutherford fellowships ST/J00412X/1, ST/M005798/1), United Kingdom;
ERC (grants 268062,  336581 and  starting grant 802836 ``AxScale'');
EU Horizon 2020 (Marie Sk\l{}odowska-Curie grants 701386, 754496, 842407, 893101).
 
\end{acknowledgments}

\bibliography{bibliography}

\onecolumngrid
{\begin{center}
NA62 Collaboration
\end{center}
%
%%%%%%%%%%%%%%%%%%%%%%%%%%%%%%%%%
%

%%%%%%%
\begin{center}
 R.~Aliberti\footnotemark[5]$^,\,$\renewcommand{\thefootnote}{\alphalph{\value{footnote}}}\footnotemark[1]\renewcommand{\thefootnote}{\arabic{footnote}},
 F.~Ambrosino\footnotemark[11],
 R.~Ammendola\footnotemark[19],
 B.~Angelucci\footnotemark[31],
 A.~Antonelli\footnotemark[10],
 G.~Anzivino\footnotemark[12],
 R.~Arcidiacono\footnotemark[20]$^,\,$\renewcommand{\thefootnote}{\alphalph{\value{footnote}}}\footnotemark[2]\renewcommand{\thefootnote}{\arabic{footnote}},
 T.~Bache\footnotemark[29],
 A.~Baeva\footnotemark[24],
 D.~Baigarashev\footnotemark[24],
 M.~Barbanera\footnotemark[13],
 J.~Bernhard\footnotemark[28],
 A.~Biagioni\footnotemark[18],
 L.~Bician\footnotemark[27]$^,\,$\renewcommand{\thefootnote}{\alphalph{\value{footnote}}}\footnotemark[3]\renewcommand{\thefootnote}{\arabic{footnote}},
 C.~Biino\footnotemark[21],
 A.~Bizzeti\footnotemark[8]$^,\,$\renewcommand{\thefootnote}{\alphalph{\value{footnote}}}\footnotemark[4]\renewcommand{\thefootnote}{\arabic{footnote}},
 T.~Blazek\footnotemark[27],
 B.~Bloch-Devaux\footnotemark[20],
 V.~Bonaiuto\footnotemark[19]$^,\,$\renewcommand{\thefootnote}{\alphalph{\value{footnote}}}\footnotemark[5]\renewcommand{\thefootnote}{\arabic{footnote}},
 M.~Boretto\footnotemark[20]$^,\,$\renewcommand{\thefootnote}{\alphalph{\value{footnote}}}\footnotemark[3]\renewcommand{\thefootnote}{\arabic{footnote}},
 A. M.~Bragadireanu\footnotemark[23],
 D.~Britton\footnotemark[31],
 F.~Brizioli\footnotemark[12],
 M. B.~Brunetti\footnotemark[29]$^,\,$\renewcommand{\thefootnote}{\alphalph{\value{footnote}}}\footnotemark[6]\renewcommand{\thefootnote}{\arabic{footnote}},
 D.~Bryman\footnotemark[3]$^,\,$\renewcommand{\thefootnote}{\alphalph{\value{footnote}}}\footnotemark[7]\renewcommand{\thefootnote}{\arabic{footnote}},
 F.~Bucci\footnotemark[8],
 T.~Capussela\footnotemark[11],
 J.~Carmignani\footnotemark[33],
 A.~Ceccucci\footnotemark[28],
 P.~Cenci\footnotemark[13],
 V.~Cerny\footnotemark[27],
 C.~Cerri\footnotemark[16],
 B.~Checcucci\footnotemark[13],
 A.~Conovaloff\footnotemark[34],
 P.~Cooper\footnotemark[34],
 E.~Cortina Gil\footnotemark[1],
 M.~Corvino\footnotemark[11]$^,\,$\renewcommand{\thefootnote}{\alphalph{\value{footnote}}}\footnotemark[3]\renewcommand{\thefootnote}{\arabic{footnote}},
 F.~Costantini\footnotemark[14],
 A.~Cotta Ramusino\footnotemark[6],
 D.~Coward\footnotemark[34]$^,\,$\renewcommand{\thefootnote}{\alphalph{\value{footnote}}}\footnotemark[8]\renewcommand{\thefootnote}{\arabic{footnote}},
 G.~D'Agostini\footnotemark[17],
 J. B.~Dainton\footnotemark[33],
 P.~Dalpiaz\footnotemark[7],
 H.~Danielsson\footnotemark[28],
 N.~De Simone\footnotemark[28]$^,\,$\renewcommand{\thefootnote}{\alphalph{\value{footnote}}}\footnotemark[9]\renewcommand{\thefootnote}{\arabic{footnote}},
 D.~Di Filippo\footnotemark[11],
 L.~Di Lella\footnotemark[14]$^,\,$\renewcommand{\thefootnote}{\alphalph{\value{footnote}}}\footnotemark[10]\renewcommand{\thefootnote}{\arabic{footnote}},
 N.~Doble\footnotemark[14]$^,\,$\renewcommand{\thefootnote}{\alphalph{\value{footnote}}}\footnotemark[10]\renewcommand{\thefootnote}{\arabic{footnote}},
 V.~Duk\footnotemark[29]$^,\,$\renewcommand{\thefootnote}{\alphalph{\value{footnote}}}\footnotemark[11]\renewcommand{\thefootnote}{\arabic{footnote}},
 F.~Duval\footnotemark[28],
 B.~D\"obrich\footnotemark[28],
 D.~Emelyanov\footnotemark[24],
 J.~Engelfried\footnotemark[22],
 T.~Enik\footnotemark[24],
 N.~Estrada-Tristan\footnotemark[22]$^,\,$\renewcommand{\thefootnote}{\alphalph{\value{footnote}}}\footnotemark[12]\renewcommand{\thefootnote}{\arabic{footnote}},
 V.~Falaleev\footnotemark[24],
 R.~Fantechi\footnotemark[16],
 V.~Fascianelli\footnotemark[29]$^,\,$\renewcommand{\thefootnote}{\alphalph{\value{footnote}}}\footnotemark[13]\renewcommand{\thefootnote}{\arabic{footnote}},
 L.~Federici\footnotemark[28],
 S.~Fedotov\footnotemark[25],
 A.~Filippi\footnotemark[21],
 M.~Fiorini\footnotemark[7],
 J. R.~Fry\footnotemark[29],
 J.~Fu\footnotemark[3],
 A.~Fucci\footnotemark[19],
 L.~Fulton\footnotemark[32],
 E.~Gamberini\footnotemark[28],
 L.~Gatignon\footnotemark[28]$^,\,$\renewcommand{\thefootnote}{\alphalph{\value{footnote}}}\footnotemark[14]\renewcommand{\thefootnote}{\arabic{footnote}},
 G.~Georgiev\footnotemark[10]$^,\,$\renewcommand{\thefootnote}{\alphalph{\value{footnote}}}\footnotemark[15]\renewcommand{\thefootnote}{\arabic{footnote}},
 S. A.~Ghinescu\footnotemark[23],
 A.~Gianoli\footnotemark[6],
 M.~Giorgi\footnotemark[14],
 S.~Giudici\footnotemark[14],
 F.~Gonnella\footnotemark[29],
 E.~Goudzovski\footnotemark[29],
 C.~Graham\footnotemark[31],
 R.~Guida\footnotemark[28],
 E.~Gushchin\footnotemark[25],
 F.~Hahn\footnotemark[28]$^,\,$\renewcommand{\thefootnote}{\fnsymbol{footnote}}\footnotemark[2]\renewcommand{\thefootnote}{\arabic{footnote}},
 H.~Heath\footnotemark[30],
 J.~Henshaw\footnotemark[29],
 E. B.~Holzer\footnotemark[28],
 T.~Husek\footnotemark[4]$^,\,$\renewcommand{\thefootnote}{\alphalph{\value{footnote}}}\footnotemark[16]\renewcommand{\thefootnote}{\arabic{footnote}},
 O. E.~Hutanu\footnotemark[23],
 D.~Hutchcroft\footnotemark[32],
 L.~Iacobuzio\footnotemark[29],
 E.~Iacopini\footnotemark[9],
 E.~Imbergamo\footnotemark[12],
 B.~Jenninger\footnotemark[28],
 J.~Jerhot\footnotemark[4]$^,\,$\renewcommand{\thefootnote}{\alphalph{\value{footnote}}}\footnotemark[17]\renewcommand{\thefootnote}{\arabic{footnote}},
 R. W. L.~Jones\footnotemark[33],
 K.~Kampf\footnotemark[4],
 V.~Kekelidze\footnotemark[24],
 S.~Kholodenko\footnotemark[26],
 G.~Khoriauli\footnotemark[5]$^,\,$\renewcommand{\thefootnote}{\alphalph{\value{footnote}}}\footnotemark[18]\renewcommand{\thefootnote}{\arabic{footnote}},
 A.~Khotyantsev\footnotemark[25],
 A.~Kleimenova\footnotemark[1],
 A.~Korotkova\footnotemark[24],
 M.~Koval\footnotemark[28]$^,\,$\renewcommand{\thefootnote}{\alphalph{\value{footnote}}}\footnotemark[19]\renewcommand{\thefootnote}{\arabic{footnote}},
 V.~Kozhuharov\footnotemark[10]$^,\,$\renewcommand{\thefootnote}{\alphalph{\value{footnote}}}\footnotemark[15]\renewcommand{\thefootnote}{\arabic{footnote}},
 Z.~Kucerova\footnotemark[27],
 Y.~Kudenko\footnotemark[25]$^,\,$\renewcommand{\thefootnote}{\alphalph{\value{footnote}}}\footnotemark[20]\renewcommand{\thefootnote}{\arabic{footnote}},
 J.~Kunze\footnotemark[5],
 V.~Kurochka\footnotemark[25],
 V.~Kurshetsov\footnotemark[26],
 G.~Lamanna\footnotemark[14],
 G.~Lanfranchi\footnotemark[10],
 E.~Lari\footnotemark[14],
 G.~Latino\footnotemark[9],
 P.~Laycock\footnotemark[28]$^,\,$\renewcommand{\thefootnote}{\alphalph{\value{footnote}}}\footnotemark[21]\renewcommand{\thefootnote}{\arabic{footnote}},
 C.~Lazzeroni\footnotemark[29],
 G.~Lehmann Miotto\footnotemark[28],
 M.~Lenti\footnotemark[9],
 E.~Leonardi\footnotemark[18],
 P.~Lichard\footnotemark[28],
 L.~Litov\footnotemark[24]$^,\,$\renewcommand{\thefootnote}{\alphalph{\value{footnote}}}\footnotemark[15]\renewcommand{\thefootnote}{\arabic{footnote}},
 R.~Lollini\footnotemark[12],
 D.~Lomidze\footnotemark[5]$^,\,$\renewcommand{\thefootnote}{\alphalph{\value{footnote}}}\footnotemark[22]\renewcommand{\thefootnote}{\arabic{footnote}},
 A.~Lonardo\footnotemark[18],
 P.~Lubrano\footnotemark[13],
 M.~Lupi\footnotemark[13]$^,\,$\renewcommand{\thefootnote}{\alphalph{\value{footnote}}}\footnotemark[23]\renewcommand{\thefootnote}{\arabic{footnote}},
 N.~Lurkin\footnotemark[29]$^,\,$\renewcommand{\thefootnote}{\alphalph{\value{footnote}}}\footnotemark[17]\renewcommand{\thefootnote}{\arabic{footnote}},
 D.~Madigozhin\footnotemark[24],
 I.~Mannelli\footnotemark[15],
 A.~Mapelli\footnotemark[28],
 F.~Marchetto\footnotemark[21],
 R.~Marchevski\footnotemark[28]$^,\,$\renewcommand{\thefootnote}{\alphalph{\value{footnote}}}\footnotemark[24]\renewcommand{\thefootnote}{\arabic{footnote}},
 S.~Martellotti\footnotemark[10],
 P.~Massarotti\footnotemark[11],
 K.~Massri\footnotemark[28],
 E.~Maurice\footnotemark[32]$^,\,$\renewcommand{\thefootnote}{\alphalph{\value{footnote}}}\footnotemark[25]\renewcommand{\thefootnote}{\arabic{footnote}},
 M.~Medvedeva\footnotemark[25],
 A.~Mefodev\footnotemark[25],
 E.~Menichetti\footnotemark[20],
 E.~Migliore\footnotemark[20],
 E.~Minucci\footnotemark[1]$^,\,$\renewcommand{\thefootnote}{\alphalph{\value{footnote}}}\footnotemark[3]\renewcommand{\thefootnote}{\arabic{footnote}}$^,\,$\renewcommand{\thefootnote}{\alphalph{\value{footnote}}}\footnotemark[26]\renewcommand{\thefootnote}{\arabic{footnote}}$^,\,$\renewcommand{\thefootnote}{\fnsymbol{footnote}}\footnotemark[1]\renewcommand{\thefootnote}{\arabic{footnote}},
 M.~Mirra\footnotemark[11],
 M.~Misheva\footnotemark[24]$^,\,$\renewcommand{\thefootnote}{\alphalph{\value{footnote}}}\footnotemark[27]\renewcommand{\thefootnote}{\arabic{footnote}},
 N.~Molokanova\footnotemark[24],
 M.~Moulson\footnotemark[10],
 S.~Movchan\footnotemark[24],
 M.~Napolitano\footnotemark[11],
 I.~Neri\footnotemark[7],
 F.~Newson\footnotemark[29],
 A.~Norton\footnotemark[7]$^,\,$\renewcommand{\thefootnote}{\alphalph{\value{footnote}}}\footnotemark[28]\renewcommand{\thefootnote}{\arabic{footnote}},
 M.~Noy\footnotemark[28],
 T.~Numao\footnotemark[2],
 V.~Obraztsov\footnotemark[26],
 A.~Ostankov\footnotemark[26]$^,\,$\renewcommand{\thefootnote}{\fnsymbol{footnote}}\footnotemark[2]\renewcommand{\thefootnote}{\arabic{footnote}},
 S.~Padolski\footnotemark[1]$^,\,$\renewcommand{\thefootnote}{\alphalph{\value{footnote}}}\footnotemark[21]\renewcommand{\thefootnote}{\arabic{footnote}},
 R.~Page\footnotemark[30],
 V.~Palladino\footnotemark[28]$^,\,$\renewcommand{\thefootnote}{\alphalph{\value{footnote}}}\footnotemark[29]\renewcommand{\thefootnote}{\arabic{footnote}},
 A.~Parenti\footnotemark[9],
 C.~Parkinson\footnotemark[29]$^,\,$\renewcommand{\thefootnote}{\alphalph{\value{footnote}}}\footnotemark[17]\renewcommand{\thefootnote}{\arabic{footnote}},
 E.~Pedreschi\footnotemark[14],
 M.~Pepe\footnotemark[13],
 M.~Perrin-Terrin\footnotemark[28]$^,\,$\renewcommand{\thefootnote}{\alphalph{\value{footnote}}}\footnotemark[30]\renewcommand{\thefootnote}{\arabic{footnote}}$^,\,$\renewcommand{\thefootnote}{\alphalph{\value{footnote}}}\footnotemark[31]\renewcommand{\thefootnote}{\arabic{footnote}},
 L.~Peruzzo\footnotemark[5],
 P.~Petrov\footnotemark[1],
 Y.~Petrov\footnotemark[2],
 F.~Petrucci\footnotemark[7],
 R.~Piandani\footnotemark[12]$^,\,$\renewcommand{\thefootnote}{\alphalph{\value{footnote}}}\footnotemark[32]\renewcommand{\thefootnote}{\arabic{footnote}},
 M.~Piccini\footnotemark[13],
 J.~Pinzino\footnotemark[28]$^,\,$\renewcommand{\thefootnote}{\alphalph{\value{footnote}}}\footnotemark[33]\renewcommand{\thefootnote}{\arabic{footnote}},
 I.~Polenkevich\footnotemark[24],
 L.~Pontisso\footnotemark[16],
 Yu.~Potrebenikov\footnotemark[24],
 D.~Protopopescu\footnotemark[31],
 M.~Raggi\footnotemark[17],
 A.~Romano\footnotemark[29],
 P.~Rubin\footnotemark[34],
 G.~Ruggiero\footnotemark[33]$^,\,$\renewcommand{\thefootnote}{\alphalph{\value{footnote}}}\footnotemark[24]\renewcommand{\thefootnote}{\arabic{footnote}},
 V.~Ryjov\footnotemark[28],
 A.~Salamon\footnotemark[19],
 C.~Santoni\footnotemark[12],
 G.~Saracino\footnotemark[11],
 F.~Sargeni\footnotemark[19]$^,\,$\renewcommand{\thefootnote}{\alphalph{\value{footnote}}}\footnotemark[34]\renewcommand{\thefootnote}{\arabic{footnote}},
 S.~Schuchmann\footnotemark[28]$^,\,$\renewcommand{\thefootnote}{\alphalph{\value{footnote}}}\footnotemark[10]\renewcommand{\thefootnote}{\arabic{footnote}},
 V.~Semenov\footnotemark[26]$^,\,$\renewcommand{\thefootnote}{\fnsymbol{footnote}}\footnotemark[2]\renewcommand{\thefootnote}{\arabic{footnote}},
 A.~Sergi\footnotemark[29]$^,\,$\renewcommand{\thefootnote}{\alphalph{\value{footnote}}}\footnotemark[35]\renewcommand{\thefootnote}{\arabic{footnote}},
 A.~Shaikhiev\footnotemark[1]$^,\,$\renewcommand{\thefootnote}{\alphalph{\value{footnote}}}\footnotemark[36]\renewcommand{\thefootnote}{\arabic{footnote}},
 S.~Shkarovskiy\footnotemark[24],
 D.~Soldi\footnotemark[20],
 M.~Sozzi\footnotemark[14],
 T.~Spadaro\footnotemark[10],
 F.~Spinella\footnotemark[16],
 A.~Sturgess\footnotemark[29],
 V.~Sugonyaev\footnotemark[26],
 J.~Swallow\footnotemark[29]$^,\,$\renewcommand{\thefootnote}{\fnsymbol{footnote}}\footnotemark[1]\renewcommand{\thefootnote}{\arabic{footnote}},
 S.~Trilov\footnotemark[30],
 P.~Valente\footnotemark[18],
 B.~Velghe\footnotemark[2],
 S.~Venditti\footnotemark[28],
 P.~Vicini\footnotemark[18],
 R.~Volpe\footnotemark[1]$^,\,$\renewcommand{\thefootnote}{\alphalph{\value{footnote}}}\footnotemark[37]\renewcommand{\thefootnote}{\arabic{footnote}},
 M.~Vormstein\footnotemark[5],
 H.~Wahl\footnotemark[7]$^,\,$\renewcommand{\thefootnote}{\alphalph{\value{footnote}}}\footnotemark[10]\renewcommand{\thefootnote}{\arabic{footnote}},
 R.~Wanke\footnotemark[5],
 B.~Wrona\footnotemark[32],
 O.~Yushchenko\footnotemark[26],
 M.~Zamkovsky\footnotemark[4],
 A.~Zinchenko\footnotemark[24]$^,\,$\renewcommand{\thefootnote}{\fnsymbol{footnote}}\footnotemark[2]\renewcommand{\thefootnote}{\arabic{footnote}}
\end{center}

%
%%%%%%%%%%%%%%%%%%%%%%%%%%%%%%%%%
%

\setcounter{footnote}{0}
\newlength{\basefootnotesep}
\setlength{\basefootnotesep}{\footnotesep}

\begin{center}
$^{1}${\centering Universit\'e Catholique de Louvain, B-1348 Louvain-La-Neuve, Belgium} \\
$^{2}${\centering TRIUMF, Vancouver, British Columbia, V6T 2A3, Canada} \\
$^{3}${\centering University of British Columbia, Vancouver, British Columbia, V6T 1Z4, Canada} \\
$^{4}${\centering Charles University, 116 36 Prague 1, Czech Republic} \\
$^{5}${\centering Institut f\"ur Physik and PRISMA Cluster of Excellence, Universit\"at Mainz, D-55099 Mainz, Germany} \\
$^{6}${\centering INFN, Sezione di Ferrara, I-44122 Ferrara, Italy} \\
$^{7}${\centering Dipartimento di Fisica e Scienze della Terra dell'Universit\`a e INFN, Sezione di Ferrara, I-44122 Ferrara, Italy} \\
$^{8}${\centering INFN, Sezione di Firenze, I-50019 Sesto Fiorentino, Italy} \\
$^{9}${\centering Dipartimento di Fisica e Astronomia dell'Universit\`a e INFN, Sezione di Firenze, I-50019 Sesto Fiorentino, Italy} \\
$^{10}${\centering Laboratori Nazionali di Frascati, I-00044 Frascati, Italy} \\
$^{11}${\centering Dipartimento di Fisica ``Ettore Pancini'' e INFN, Sezione di Napoli, I-80126 Napoli, Italy} \\
$^{12}${\centering Dipartimento di Fisica e Geologia dell'Universit\`a e INFN, Sezione di Perugia, I-06100 Perugia, Italy} \\
$^{13}${\centering INFN, Sezione di Perugia, I-06100 Perugia, Italy} \\
$^{14}${\centering Dipartimento di Fisica dell'Universit\`a e INFN, Sezione di Pisa, I-56100 Pisa, Italy} \\
$^{15}${\centering Scuola Normale Superiore e INFN, Sezione di Pisa, I-56100 Pisa, Italy} \\
$^{16}${\centering INFN, Sezione di Pisa, I-56100 Pisa, Italy} \\
$^{17}${\centering Dipartimento di Fisica, Sapienza Universit\`a di Roma e INFN, Sezione di Roma I, I-00185 Roma, Italy} \\
$^{18}${\centering INFN, Sezione di Roma I, I-00185 Roma, Italy} \\
$^{19}${\centering INFN, Sezione di Roma Tor Vergata, I-00133 Roma, Italy} \\
$^{20}${\centering Dipartimento di Fisica dell'Universit\`a e INFN, Sezione di Torino, I-10125 Torino, Italy} \\
$^{21}${\centering INFN, Sezione di Torino, I-10125 Torino, Italy} \\
$^{22}${\centering Instituto de F\'isica, Universidad Aut\'onoma de San Luis Potos\'i, 78240 San Luis Potos\'i, Mexico} \\
$^{23}${\centering Horia Hulubei National Institute of Physics for R\&D in Physics and Nuclear Engineering, 077125 Bucharest-Magurele, Romania} \\
$^{24}${\centering Joint Institute for Nuclear Research, 141980 Dubna (MO), Russia} \\
$^{25}${\centering Institute for Nuclear Research of the Russian Academy of Sciences, 117312 Moscow, Russia} \\
$^{26}${\centering Institute for High Energy Physics - State Research Center of Russian Federation, 142281 Protvino (MO), Russia} \\
$^{27}${\centering Faculty of Mathematics, Physics and Informatics, Comenius University, 842 48, Bratislava, Slovakia} \\
$^{28}${\centering CERN,  European Organization for Nuclear Research, CH-1211 Geneva 23, Switzerland} \\
$^{29}${\centering University of Birmingham, Edgbaston, Birmingham, B15 2TT, UK} \\
$^{30}${\centering University of Bristol, Bristol, BS8 1TH, UK} \\
$^{31}${\centering University of Glasgow, Glasgow, G12 8QQ, UK} \\
$^{32}${\centering University of Liverpool, Liverpool, L69 7ZE, UK} \\
$^{33}${\centering University of Lancaster, Lancaster, LA1 4YW, UK} \\
$^{34}${\centering George Mason University, Fairfax, VA 22030, USA} \\
\end{center}
\vspace{10pt}
\renewcommand{\thefootnote}{\fnsymbol{footnote}}
$^{\footnotemark[1]}${elisa.minucci@cern.ch,  joel.christopher.swallow@cern.ch}\\
$^{\footnotemark[2]}${Deceased}
\vspace{10pt} \\
\renewcommand{\thefootnote}{\alphalph{\value{footnote}}}
$^{\alphalph{1}}${Present address: Institut f\"ur Kernphysik and Helmholtz Institute Mainz, Universit\"at Mainz, Mainz, D-55099, Germany} \\
$^{\alphalph{2}}${Also at Universit\`a degli Studi del Piemonte Orientale, I-13100 Vercelli, Italy} \\
$^{\alphalph{3}}${Present address: CERN,  European Organization for Nuclear Research, CH-1211 Geneva 23, Switzerland} \\
$^{\alphalph{4}}${Also at Dipartimento di Fisica, Universit\`a di Modena e Reggio Emilia, I-41125 Modena, Italy} \\
$^{\alphalph{5}}${Also at Department of Industrial Engineering, University of Roma Tor Vergata, I-00173 Roma, Italy} \\
$^{\alphalph{6}}${Present address: Department of Physics, University of Warwick, Coventry, CV4 7AL, UK} \\
$^{\alphalph{7}}${Also at TRIUMF, Vancouver, British Columbia, V6T 2A3, Canada} \\
$^{\alphalph{8}}${Also at SLAC National Accelerator Laboratory, Stanford University, Menlo Park, CA 94025, USA} \\
$^{\alphalph{9}}${Present address: DESY, D-15738 Zeuthen, Germany} \\
$^{\alphalph{10}}${Present address: Institut f\"ur Physik and PRISMA Cluster of Excellence, Universit\"at Mainz, D-55099 Mainz, Germany} \\
$^{\alphalph{11}}${Present address: INFN, Sezione di Perugia, I-06100 Perugia, Italy} \\
$^{\alphalph{12}}${Also at Universidad de Guanajuato, 36000 Guanajuato, Mexico} \\
$^{\alphalph{13}}${Present address: Center for theoretical neuroscience, Columbia University, New York, NY 10027, USA} \\
$^{\alphalph{14}}${Present address: University of Lancaster, Lancaster, LA1 4YW, UK} \\
$^{\alphalph{15}}${Also at Faculty of Physics, University of Sofia, BG-1164 Sofia, Bulgaria} \\
$^{\alphalph{16}}${Present address: Department of Astronomy and Theoretical Physics, Lund University, Lund, SE 223-62, Sweden} \\
$^{\alphalph{17}}${Present address: Universit\'e Catholique de Louvain, B-1348 Louvain-La-Neuve, Belgium} \\
$^{\alphalph{18}}${Present address: Universit\"at W\"urzburg, D-97070 W\"urzburg, Germany} \\
$^{\alphalph{19}}${Present address: Charles University, 116 36 Prague 1, Czech Republic} \\
$^{\alphalph{20}}${Also at National Research Nuclear University (MEPhI), 115409 Moscow and Moscow Institute of Physics and Technology, 141701 Moscow region, Moscow, Russia} \\
$^{\alphalph{21}}${Present address: Brookhaven National Laboratory, Upton, NY 11973, USA} \\
$^{\alphalph{22}}${Present address: European XFEL GmbH, D-22761 Hamburg, Germany} \\
$^{\alphalph{23}}${Present address: Institut am Fachbereich Informatik und Mathematik, Goethe Universit\"at, D-60323 Frankfurt am Main, Germany} \\
$^{\alphalph{24}}${Present address: Dipartimento di Fisica e Astronomia dell'Universit\`a e INFN, Sezione di Firenze, I-50019 Sesto Fiorentino, Italy} \\
$^{\alphalph{25}}${Present address: Laboratoire Leprince Ringuet, F-91120 Palaiseau, France} \\
$^{\alphalph{26}}${Also at Laboratori Nazionali di Frascati, I-00044 Frascati, Italy} \\
$^{\alphalph{27}}${Present address: Institute of Nuclear Research and Nuclear Energy of Bulgarian Academy of Science (INRNE-BAS), BG-1784 Sofia, Bulgaria} \\
$^{\alphalph{28}}${Present address: University of Glasgow, Glasgow, G12 8QQ, UK} \\
$^{\alphalph{29}}${Present address: Physics Department, Imperial College London, London, SW7 2BW, UK} \\
$^{\alphalph{30}}${Present address: Aix Marseille University, CNRS/IN2P3, CPPM, F-13288, Marseille, France} \\
$^{\alphalph{31}}${Also at Universit\'e Catholique de Louvain, B-1348 Louvain-La-Neuve, Belgium} \\
$^{\alphalph{32}}${Present address: University of Chinese Academy of Sciences, Beijing, 100049, China} \\
$^{\alphalph{33}}${Present address: INFN, Sezione di Pisa, I-56100 Pisa, Italy} \\
$^{\alphalph{34}}${Also at Department of Electronic Engineering, University of Roma Tor Vergata, I-00173 Roma, Italy} \\
$^{\alphalph{35}}${Present address: Dipartimento di Fisica dell'Universit\`a e INFN, Sezione di Genova, I-16146 Genova, Italy} \\
$^{\alphalph{36}}${Also at Institute for Nuclear Research of the Russian Academy of Sciences, 117312 Moscow, Russia} \\
$^{\alphalph{37}}${Present address: Faculty of Mathematics, Physics and Informatics, Comenius University, 842 48, Bratislava, Slovakia} \\
\renewcommand{\thefootnote}{\arabic{footnote}}

\end{document}